\documentclass[aps,prc,twocolumn,floatfix,showpacs,nofootinbib,amsmath,amssymb]{revtex4-1}

\usepackage{graphicx}
\usepackage{dcolumn}
\usepackage{bm}
\usepackage{color}

\newcommand{\be}{\begin{equation}}
\newcommand{\ee}{\end{equation}}
\newcommand{\ba}{\begin{eqnarray}}
\newcommand{\ea}{\end{eqnarray}}
\newcommand{\bd}{\begin{displaymath}}
\newcommand{\ed}{\end{displaymath}}
\newcommand{\bea}{\begin{eqnarray}}
\newcommand{\eea}{\end{eqnarray}}

\begin{document}
\title{Embedding a critical point in a hadron to quark--gluon crossover equation of state}
\author{J. I. Kapusta}
\affiliation{School of Physics \& Astronomy, University of Minnesota, Minneapolis, MN 55455, USA}
\author{C. Plumberg}
\affiliation{Illinois Center for Advanced Studies of the Universe, Department of Physics, University of Illinois at Urbana-Champaign, Urbana, IL 61801, USA}
\author{T. Welle}
\affiliation{School of Physics \& Astronomy, University of Minnesota, Minneapolis, MN 55455, USA}

\begin{abstract}
Lattice QCD simulations have shown unequivocally that the transition from hadrons to quarks and gluons is a crossover when the baryon chemical potential is zero or small.  Many model calculations predict the existence of a critical point at a value of the chemical potential where current lattice simulations are unreliable.  We show how to embed a critical point in a smooth background equation of state so as to yield the critical exponents and critical amplitude ratios expected of a transition in the same universality class as the liquid--gas phase transition and the 3D Ising model.  The resulting equation of state has parameters which may be inferred by hydrodynamic modeling of heavy ion collisions in the Beam Energy Scan II at the Relativistic Heavy Ion Collider or in experiments at other accelerators.
\end{abstract}

\date{\today}

\maketitle


The QCD (Quantum Chromodynamics) equation of state has been a subject of intense interest ever since the discovery of asymptotic freedom.  At high temperature $T$ and baryon chemical potential $\mu$ it is a weakly interacting gas of quarks and gluons, while at low $T$ and $\mu$ it is a strongly interacting gas of hadrons.  Lattice QCD simulations have shown conclusively that the transition from one phase to the other at $T \approx 155$ MeV and $\mu = 0$ is smooth on account of the fact that the up and down quark masses, and consequently the pion mass, are not zero.  However, diverse model calculations predict the existence of a line of first order phase transition, beginning at $T=0$ and $\mu_0$ and terminating in a critical point at $T_c < 155$ MeV and $\mu_c < \mu_0$ \cite{stephanov,CPOD}.   Such a purported critical point appears to be beyond the reach of reliable lattice calculations.  Experiments during the Beam Energy Scan II at RHIC (Relativistic Heavy Ion Collider) may or may not support the existence of critical behavior \cite{QMseries}.  The goal of this paper is to propose a general construction for the equation of state which is consistent with (i) lattice QCD for all $T$ and small $\mu$, (ii) perturbative QCD for large $T$ and/or large $\mu$, and (iii) a critical point with critical exponents and amplitude ratios from the same universality class as the liquid--gas phase transition and the 3D Ising model.  Parameters in this construction can be adjusted to best fit the experimental data.  The construction is different than the one proposed in Refs. \cite{attract,Taylor1}, which are based on the work of Refs. \cite{Zinnbook,Guida}.  These two constructions can perhaps be viewed as alternatives which provide some idea as to the range of uncertainty in how to describe matter near the critical point and the associated critical line of first order phase transition.

Without loss of generality the equation of state can be expressed as
\be
P(T,\mu) = P_{BG}(T,\mu) R(T,\mu)
\ee
where $P_{BG}(T,\mu)$ is a judiciously chosen background equation of state with no critical behavior; all critical behavior resides in the dimensionless function $R$.  Motivated by solutions to the cubic equation, which produce S-shaped curves characteristic of the van der Waals equation of state or the liquid-gas phase transition \cite{Kapusta1,Goodman}, we consider the auxilliary functions
\be
Q_{\pm}(T,\mu) = \left\{  \left[ (\Delta^2(T))^{2} + r^{2}(T,\mu) \right]^{1/2} \pm r(T,\mu)  \right\}^k
\ee
where 
\be
r(T,\mu) = \frac{\mu^m - \mu_x^m(T)}{\mu^m + \mu_x^m(T)}
\ee
with $m$ a positive even integer.  The function $\mu_x(T)$ represents the chemical potential where the two phases are in coexistence when $T \le T_c$, but it must also be a smooth function for all $T \ge T_c$ to avoid undesired discontinuities.  Note that $-1 \le r < 1$ and that it vanishes along the coexistence curve.  The function $\Delta^2(T)$ is expected to have the functional form $|T/T_c - 1|^p$ near $T_c$.  The parameters $k$ and $p$ will determine the four critical exponents.

When $T > T_c$ we take
\ba
\lefteqn{R(T,\mu) = 1 - a(T)\left( \sqrt{ \Delta^4 + 1} + 1 \right)^k} \nonumber \\
&&- a(T)\left( \sqrt{ \Delta^4 + 1} - 1 \right)^k + a(T)(Q_+ + Q_-)
\ea 
where $a(T)$ is a smooth function.  This has the property that $P \rightarrow P_{BG}$ as $\mu \rightarrow 0$ and as $\mu \rightarrow \infty$ for any fixed value of $T$.  It is an even function of $\mu$.  The density is
\be 
n(T,\mu) = n_{BG}(T,\mu) R(T,\mu) + P_{BG}(T,\mu) \frac{\partial R(T,\mu)}{\partial \mu}
\ee
The critical exponent $\delta$ is determined by $P - P_c \sim {\rm sgn}( n - n_c) |n - n_c|^\delta$ as $n \rightarrow n_c$ along the critical isotherm $\Delta^2 = 0$.  Assuming  $1 < k < 2$ the leading behavior is
\ba
n - n_c &=& \frac{m^k k a(T_c)}{\mu_c}  P_{BG}(T_c,\mu_c)
{\rm sgn}( \mu - \mu_c) \left|\frac{\mu - \mu_c}{\mu_c} \right|^{k-1} \nonumber \\
P - P_c &=& (1 - 2^k a(T_c)) n_{BG}(T_c,\mu_c) (\mu - \mu_c) 
\ea
Thus the critical exponent $\delta = 1/(k-1)$ or $k = 1 + 1/\delta$.

The baryon number susceptibility is $\chi_B = \partial^2 P/\partial \mu^2 \equiv \chi_{\mu\mu}$.  It diverges like $\chi_B = \chi_+ (T/T_c - 1)^{-\gamma}$ as 
$T \rightarrow T_c^+$ with $|\mu/\mu_c - 1| \ll T/T_c - 1$.  With $\Delta^2(T) = d_+ (T/T_c - 1)^p$ near $T_c$ we find that the susceptibility diverges as
\be
\chi_B \rightarrow \frac{m^2 k^2 a(T_c)}{2\mu_c^2} d_+^{\, k-2} P_{BG}(T_c,\mu_c) \left( \frac{T}{T_c} - 1 \right)^{-(2-k)p}
\label{chiB+}
\ee
Thus the critical exponent $\gamma = (2-k)p$ or $p = \gamma \delta/(\delta - 1)$.

The heat capacity at fixed volume is
\be
c_V = T \frac{\partial s}{\partial T}(T,n) = T \left( \chi_{TT} - \frac{\chi^2_{T\mu}}{\chi_{\mu\mu}} \right)
\label{cV}
\ee
The critical behavior is $c_V \rightarrow c_+ (T/T_c - 1)^{-\alpha}$.  It can be shown, albeit numerically, that approaching the critical point along $r=0$ gives the same result as approaching it at fixed $n_c$.  The amplitude is
\be
c_+ = 2p(2kp-k-p) d_+^k \frac{a(T_c)}{T_c} P_{BG}(T_c,\mu_c)
\ee
with $\alpha = 2-kp = 2 - \gamma (\delta + 1)/(\delta - 1)$.

When $T < T_c$ we take the pressure in the quark phase when $\mu \ge \mu_x$ to be
\be
P_Q(T,\mu) = P_{BG}(T,\mu) R_Q(T,\mu)
\ee
with
\be
R_Q = 1 + a(T) Q_+(T,\mu) - a(T) \left( \sqrt{ \Delta^4 + 1} + 1 \right)^k
\ee
and in the hadron phase when $\mu \le \mu_x$ to be
\be
P_H(T,\mu) = P_{BG}(T,\mu) R_H(T,\mu)
\ee
with
\be
R_H = 1 + a(T) Q_-(T,\mu) - a(T) \left( \sqrt{ \Delta^4 + 1} + 1 \right)^k
\ee
We refer to these as quark and hadron phases because, even though the background equation of state is a crossover, one is predominantly comprised of quarks and gluons while the other is predominantly comprised of hadrons.  Note that the pressure along the critical isotherm is
\be
P(T_c,\mu) = P_{BG}(T_c,\mu) \left[ 1 + 2^k a(T_c) \left( \left|\frac{\mu^m - \mu_c^m}{\mu^m + \mu_c^m} \right|^k -1 \right) \right]
\ee
no matter whether $T_c$ is approached from below or above.  Hence the critical exponent $\delta$ is well defined.

The density difference along the coexistence curve is
\be
\Delta n(T) = \frac{m k a(T)}{\mu_x(T)} P_{BG}(T,\mu_x(T)) \left[ \Delta^2(T) \right]^{k-1}
\label{dendiff}
\ee
The critical exponent $\beta$ is defined via $\Delta n \sim (1 - T/T_c)^\beta$.  With $\Delta^2(T) = d_- (1 - T/T_c)^p$ near $T_c$ we find that 
$\beta = p (k-1) = \gamma/(\delta - 1)$.

The susceptibility along the coexistence curve is
\ba
\chi_B(T) &=& P_{BG}(T,\mu_x) \left[ \frac{ m^2 k^2 a}{4\mu_x^2} \left( \Delta^2 \right)^{k-2}
\mp \frac{m k a}{2\mu_x^2} \left( \Delta^2 \right)^{k-1} \right] \nonumber \\
&+& \chi_{B,BG}(T,\mu_x) \left[ 1 + a(\Delta^2)^k - a\left( \sqrt{\Delta^4 + 1} + 1 \right)^k  \right] \nonumber \\
&\pm& \frac{m k a}{\mu_x} n_{BG}(T,\mu_x) \left( \Delta^2 \right)^{k-1}
\label{chiB}
\ea
Recalling that $\gamma = (2-k)p$ we write the critical part as $\chi_B(T) = \chi_- (1 - T/T_c)^{-\gamma}$.  When $T_c$ is approached from above at $n_c$ the critical part is $\chi_B(T) = \chi_+ (T/T_c - 1)^{-\gamma}$.  From the above equations the ratio of critical amplitudes is
\be
\frac{\chi_+}{\chi_-} = 2 \left( \frac{d_-}{d_+} \right)^{2-k}
\ee
Mathematically the 2 arises because above $T_c$ the sum $Q_+ + Q_-$ enters whereas below $T_c$ only $Q_+$ or $Q_-$ does.  For the universality class which includes the liquid--gas phase transition and the 3D Ising model the exact result is $\chi_+/\chi_- \approx 5$, whereas in mean field approximation $\chi_+/\chi_- = 2$.  For the latter then $d_+ = d_-$.

The last quantity to examine when $T < T_c$ is the heat capacity.  The critical behavior is $c_V(T) = c_- (1 - T/T_c)^{-\alpha}$ when the critical point is approached along the coexistence curve.  We find that
\be
c_- = p(2kp-k-p) d_-^k \frac{a(T_c)}{T_c} P_{BG}(T_c,\mu_c)
\ee
The critical exponent $\alpha$ is the same above and below $T_c$ as it should be.  The ratio of critical amplitudes is
\be
\frac{c_+}{c_-} = 2 \left( \frac{d_+}{d_-} \right)^k
\ee

This model has two independent exponents, $k$ and $p$, in terms of which the critical exponents $\alpha$, $\beta$, $\gamma$ and $\delta$ are expressed.  They obey the known relations $\alpha + 2\beta + \gamma = 2$ and $\gamma = \beta (\delta - 1)$.  The true critical exponents for the universality class which includes the liquid--gas phase transition and the 3D Ising model are $\alpha \approx 0.1101$, $\beta \approx 0.3264$, $\gamma \approx 1.2371$, and $\delta \approx 4.7898$ \cite{bootstrap1,bootstrap2} which results in $p \approx 1.564$ and $k \approx 1.209$.  In the mean field approximation $\alpha = 0$, $\beta = 1/2$, $\gamma = 1$, and $\delta = 3$ which results in $p = 3/2$ and $k = 4/3$.  In mean field approximation there is a discontinuity in $c_V$ but no divergence.  Using the true critical exponents and assuming $d_+/d_- = 1/3$ yields the ratios of critical amplitudes $c_+/c_- = 0.530$ and $\chi_+/\chi_- = 4.769$.  Given that the amplitude ratios have an uncertainty of a few percent these numbers are entirely consistent with published results \cite{Zinnbook,Guida,Hasenbusch1,Hasenbusch2}.  

For the background equation of state we choose the one described in Ref. \cite{matchingpaper}.  It uses a switching function to transition smoothly from a hadron resonance gas, with excluded volume interactions, to a perturbative quark--gluon plasma.  The switching function is
\be  
S(T,\mu) = \exp\left[ - \left(T^2/T_s^2 + \mu^2/\mu_s^2\right)^{-2} \right]
\ee
It ranges between 0 and 1 as $\mu$ and $T$ increase.  It is an even function of $\mu$ and infinitely differentiable so as not to introduce an artificial phase transition of any order.  The parameters $T_s$ and $\mu_s$ were adjusted to give a good representation of lattice results \cite{latticeQCD}, which results in $T_s = 195$ MeV and $\mu_s = 1300$ MeV.

\begin{figure}[h]
\includegraphics[width=1.07\columnwidth]{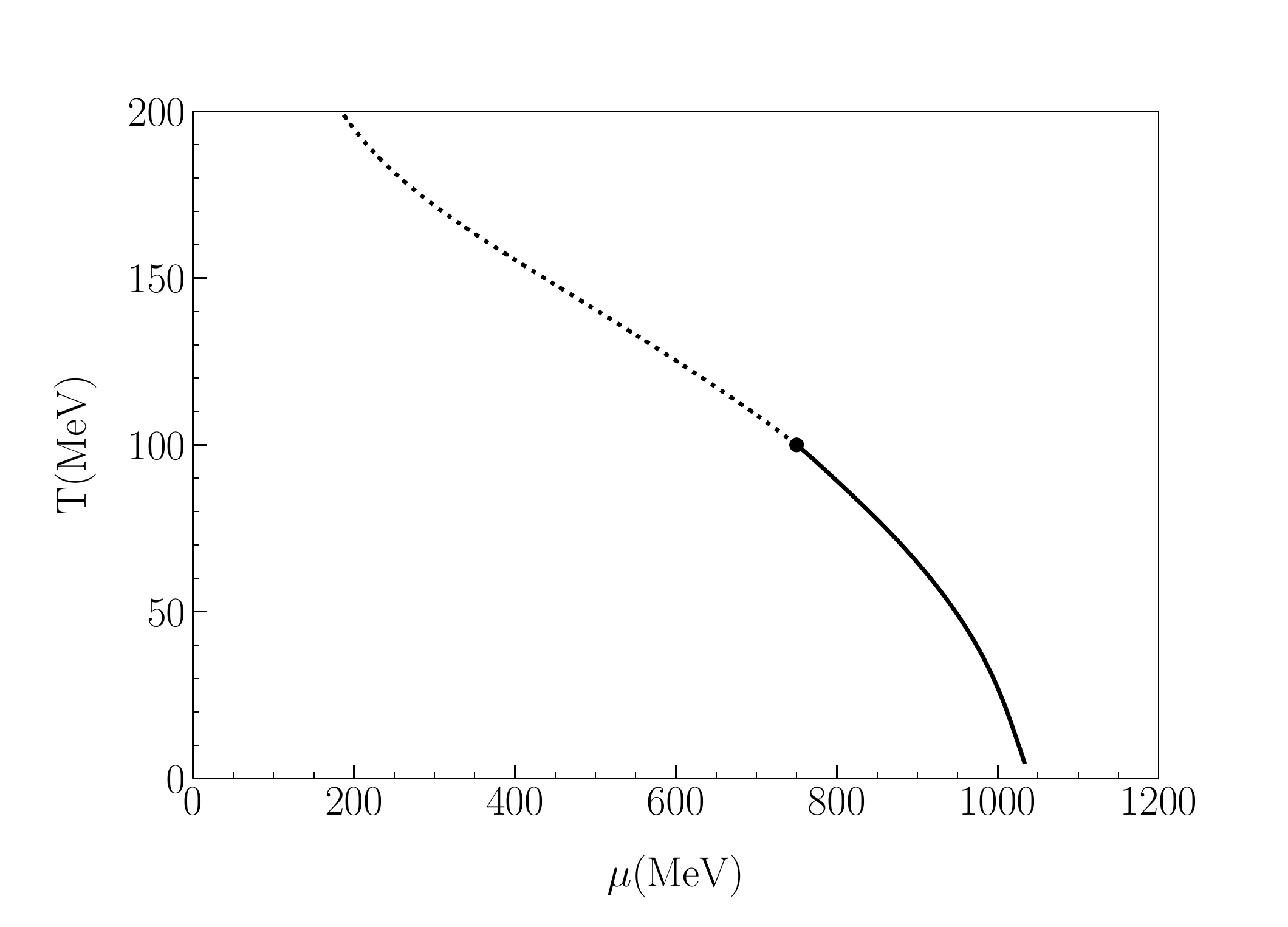}
\caption{The choice of critical curve as described in the text.  The critical temperature is taken to be 100 MeV and the critical chemical potential to be 750 MeV.}
\label{mux}
\end{figure}

Only behavior in the immediate vicinity of a critical point is universal.  For heavy ion collisions it is not even clear whether one can probe it closely enough to reveal the true critical exponents or whether mean field values are more appropriate.  Choosing the functions appearing in $R(T,\mu)$ is informed guesswork. 
Here we choose the function $\mu_x (T)$ to follow a curve of constant density. Other choices are possible, but this one produces an inverted U-shape in the $T$ versus $n$ plane.  It is determined implicitly via the relation
\be
R(T,\mu_x(T)) n_{BG}(T,\mu_x(T)) = n_c
\ee
This function is displayed in Figure \ref{mux}.  The critical point lies along this curve.  For illustration we choose $T_c = 100$ MeV and $\mu_c = 750$ MeV.

\begin{figure}[t]
\includegraphics[width=1.01\columnwidth]{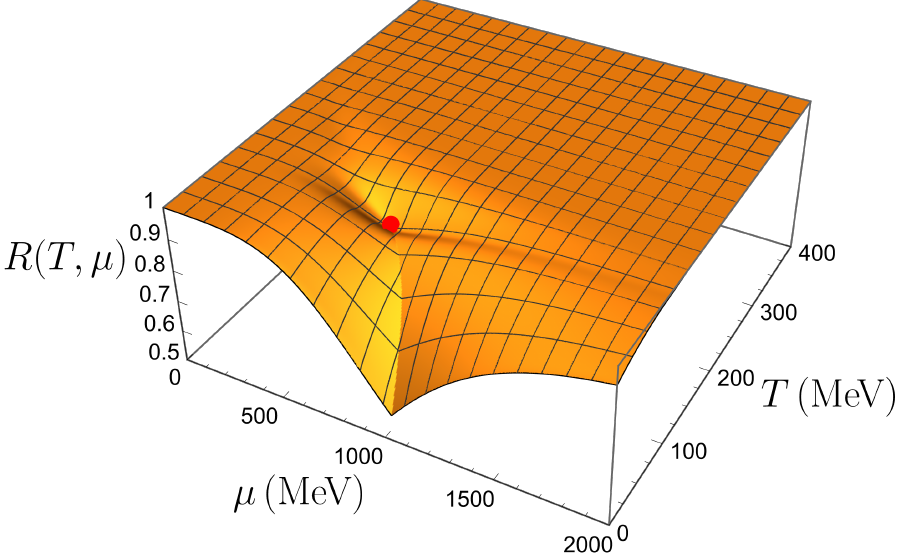}
\caption{The dimensionless function $R$ with parameters given in the text.  The critical point is indicated by the solid dot.}
\label{R}
\end{figure}

As can be seen from Eq. (\ref{dendiff}), the strength of the transition is directly proportional to the exponent $m$.  We choose $m=4$ for illustration.  The remaining functions are parameterized as
\ba
a(T) &=& a_0 \exp(-T/T_a) \nonumber \\
\Delta^2(T) &=& d_+ (T/T_c - 1)^p \exp(-T/T_d) \;\; {\rm when} \;\; T \ge T_c \nonumber \\ 
\Delta^2(T) &=& d_- (1-T/T_c)^p \exp(-T/T_d) \;\; {\rm when} \;\;  T \le T_c \nonumber \\
\ea
The parameters $T_a$ and $T_d$ determine how fast $R \rightarrow 1$ as $T$ increases beyond $T_c$.  We use the true critical exponents and amplitude ratios with parameters $d_- = 50$, $a_0 = 0.15$, $T_a = 80$ MeV, and $T_d = 200$ MeV.  Figure \ref{R} is a contour plot of the function $R(T,\mu)$.  Note that $R \rightarrow 1$ as $\mu \rightarrow 0$ and for large $T$ and/or $\mu$. 

\begin{figure}[h]
\includegraphics[width=1.07\columnwidth]{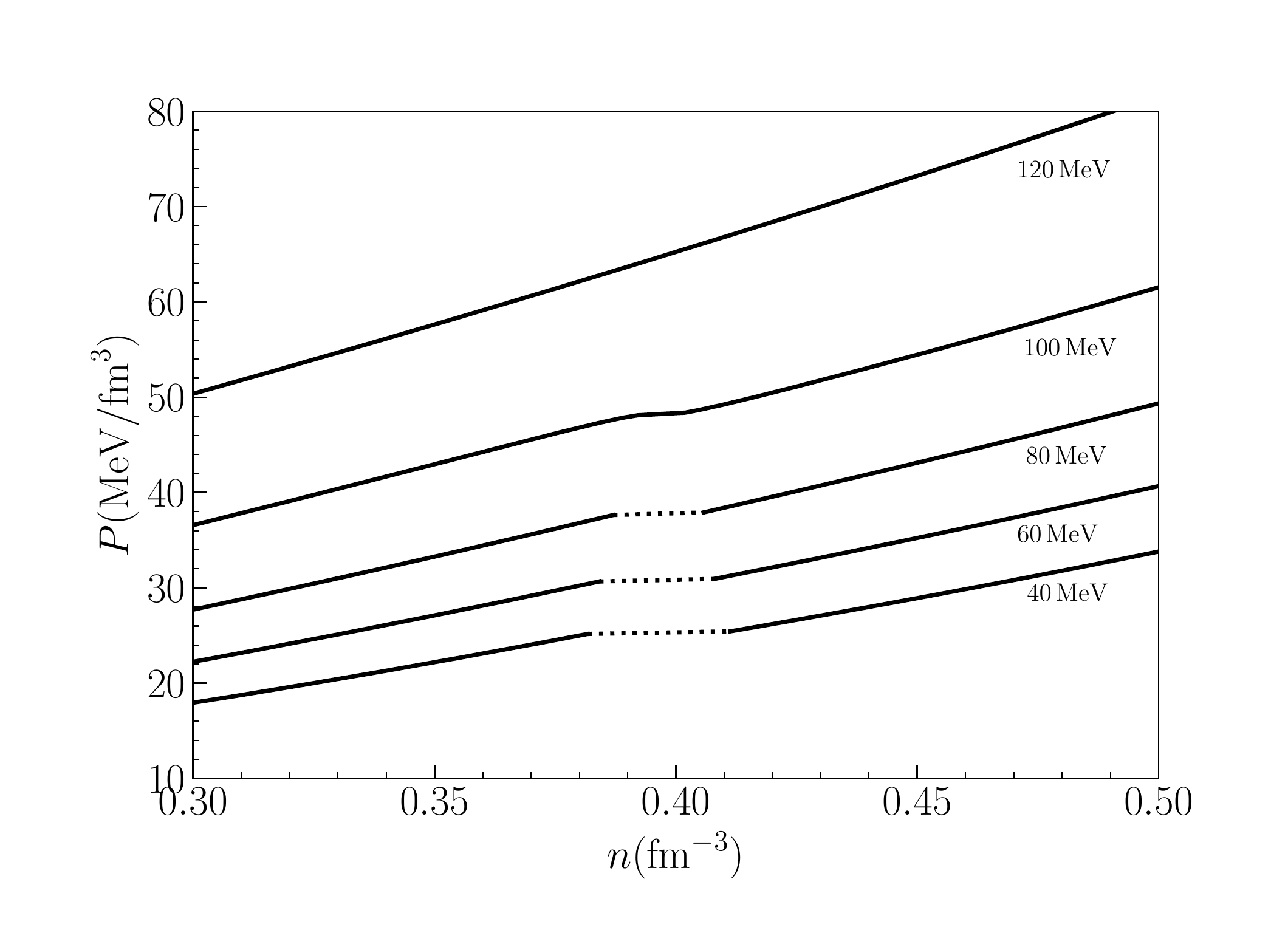}
\caption{Isotherms of pressure versus density.  The numbers label the temperature with $T_c = 100$ MeV.}
\label{P_vs_n}
\end{figure}

\begin{figure}[h]
\includegraphics[width=1.1\columnwidth]{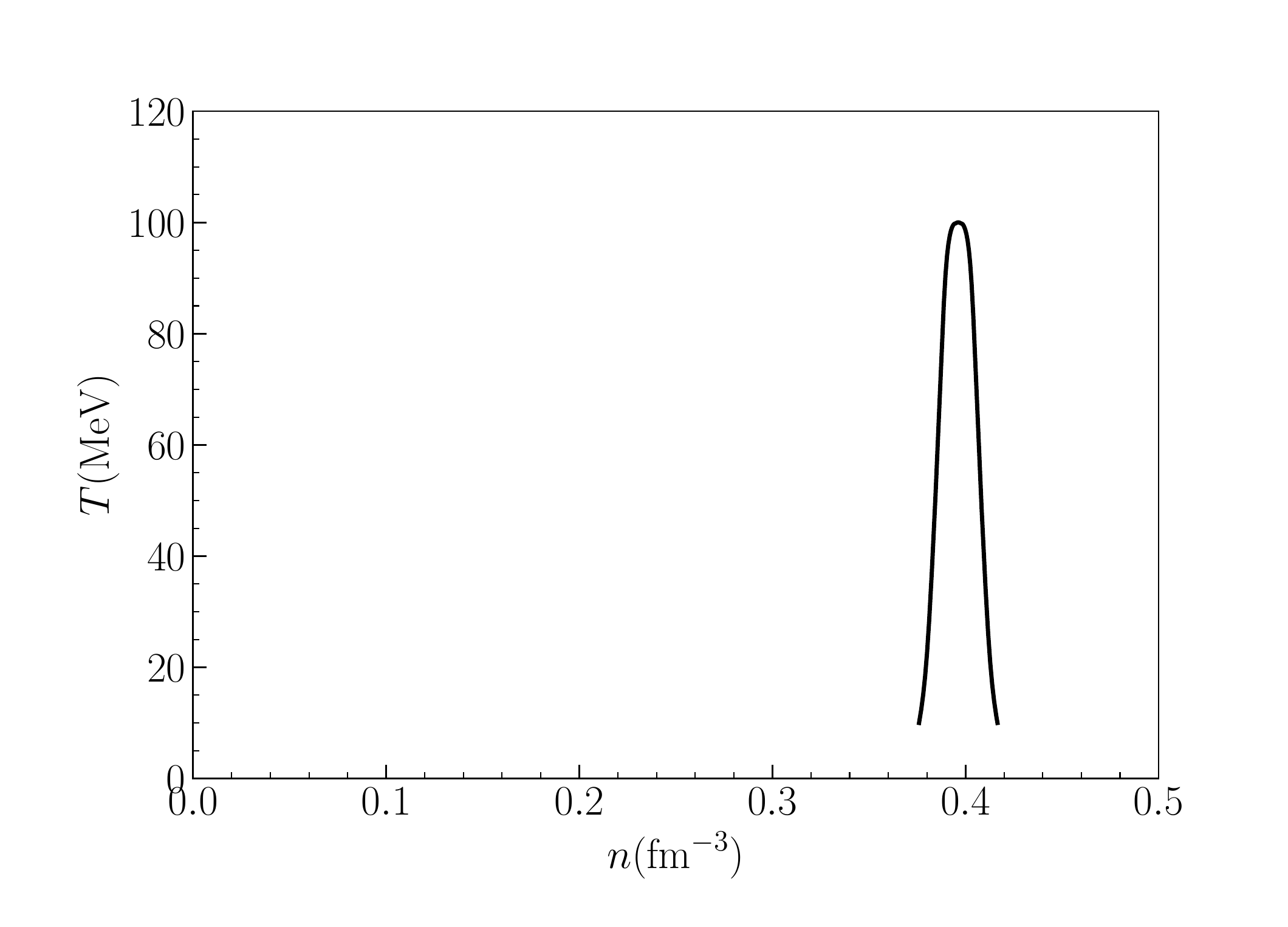}
\caption{Temperature versus density separating the two phases.}
\label{T_vs_n}
\end{figure}

Isotherms of pressure versus density are shown in Fig. \ref{P_vs_n}.  Even at $T_c$ there is almost a plateau on account of the large critical exponent 
$\delta \approx 4.79$ relative to the mean field value of 3.  
Figure \ref{T_vs_n} shows the phase transition in the temperature versus density plane.  It has the shape of an inverted U.  Other shapes are possible using different functions $\mu_x(T)$.

In conclusion, we have proposed a novel way to embed critical behavior in background equations of state which exhibit a smooth crossover from hadrons to quarks and gluons.  The goal is to use these equations of state in hydrodynamic simulations of heavy ion collisions in order to infer whether there is critical behavior.  The approach has flexibility in selecting the location of the critical point, the coexistence curve, and the reach of these into the background equation of state.  This flexibility is an advantage, as it allows parameters to be adjusted to best fit experimental data.  Examples were provided; further details and exploration, such as using a background equation of state which includes more realistic attractive and repulsive nuclear interactions at low temperature, will be published elsewhere. 

\section*{Acknowledgments}
The work of J. I. K. and T. W. was supported by the U.S. DOE Grant No. DE-FG02-87ER40328.  The work of C. P. was supported by the U.S. DOE Grant No. DE-SC0020633.


\end{document}